\newcommand{\Nside}{\text{Nside}}

\documentclass[a4paper,11pt]{article}
\usepackage{jcappub}
\usepackage{graphicx,epsfig,natbib,color,times,bm,amsmath,hyperref}
\usepackage[normalem]{ulem}
\usepackage{soul} 

\toccontinuoustrue

\begin{document}
\title{Footprints of Loop I on Cosmic Microwave Background Maps}

\author[a]{Sebastian von Hausegger,}
\author[a,b]{Hao Liu,}
\author[c]{Philipp Mertsch,}
\author[a,d]{Subir Sarkar}
\affiliation[a]{Discovery Center, Niels Bohr Institute, Blegdamsvej 17, DK-2100 Copenhagen, Denmark}
\affiliation[b]{The Key laboratory of Particle and Astrophysics,
  Institute of High Energy Physics, CAS, China}
\affiliation[c]{Kavli Institute for Particle Astrophysics \&
  Cosmology, 2575 Sand Hill Road, Menlo Park, CA 94025, USA}
\affiliation[d]{Rudolf Peierls Centre for Theoretical Physics,
  University of Oxford, 1 Keble Road, Oxford OX1 3NP, UK}
  
\emailAdd{s.vonhausegger@nbi.dk, liuhao@nbi.dk, pmertsch@stanford.edu, s.sarkar@physics.ox.ac.uk}

\abstract{Cosmology has made enormous progress through studies of the
  cosmic microwave background, however the subtle signals being now
  sought such as $B$-mode polarisation due to primordial gravitational
  waves are increasingly hard to disentangle from residual Galactic
  foregrounds in the derived CMB maps. We revisit our finding that on
  large angular scales there are traces of the nearby old supernova
  remnant Loop I in the WMAP 9-year map of the CMB and confirm this
  with the new SMICA map from the Planck satellite.}

\maketitle

\section{Introduction}
\label{sec:introduction}

We have shown earlier~\citep{Liu:2014mpa} that there is a clear imprint of the
Galactic `radio Loop~I'~\cite{Berkhuijsen:1971aa} on the internal
linear combination (ILC) map of the cosmic microwave background (CMB)
constructed using multi-frequency data obtained by the Wilkinson
Microwave Anisotropy Probe (WMAP)~\cite{Bennett:2012zja}. We
interpreted this anomaly~\citep{Liu:2014mpa} as possibly due to
magnetic dipole emission (MDE) from magnetised dust
grains~\cite{Draine:2012wr,Draine:2012zu} which can mimic a thermal
blackbody like the CMB over a restricted frequency range. 
However MDE is not an unique mechanism to explain the Loop I anomaly, rather it could be a more complex interplay between various Galactic foreground emissions (viz. free-free, thermal dust and synchrotron) which vary across the sky.

Subsequently it was claimed that our analysis had not correctly
selected the hottest peaks in the simulated sky realisations and that
it was sensitive to the map pixelization used, such that with a
different pixel size, the peak-to-loop angular distance for the
highest temperature bins is no longer
unusual~\cite{Ogburn:2014eca}. Consideration of these effects was
stated to reduce the significance of the Loop~I--CMB correlation such
that the chance probability is as large as
$\sim1\%$~\cite{Ogburn:2014eca}, rather than 0.018\%
\citep{Liu:2014mpa}.

In their recent study of low frequency Galactic foregrounds, the
Planck Collaboration~\citep{Ade:2015qkp} add the further concern that
the physical structure of Loop~I may differ~\cite{Vidal:2014boa} from
our modelling of it as a spherical shell of uniform emissivity
\cite{Berkhuijsen:1971aa,Mertsch:2013pua}. They also argue against the
anomalous correlated signal being MDE on the grounds that since there
should be many more such old supernova remnants (SNRs) in the Galactic
plane~\cite{Mertsch:2013pua} but no such signal is apparent in the
WMAP ILC map.

We demonstrate here with a new statistic which measures the clustering
of hot spots along Loop~I that our previous significance estimate is
in fact \emph{robust}. We show in addition that the correlation between Loop
I and inferred maps of the CMB holds also for the Planck SMICA
map~\citep{Adam:2015tpy} and that this is not sensitive to any
Galactic mask used. We address the second criticism by presenting
additional analyses and explanations below.

\section{Mean temperature around Loop~I}
\label{sec:Tmean}

As reported for the WMAP 9\,yr ILC
map~\cite{Bennett:2012zja}, the average temperature around the Loop~I
ring is anomalously high and the probability of finding an average
temperature at least as high in a $\Lambda$CDM random Gaussian field
is $\sim1\%$~\cite{Liu:2014mpa}. We repeat the analysis for the 2015 Planck SMICA
map~\cite{Adam:2015tpy} and investigate the effect of masking the
Galactic plane. (Hereafter we refer to these maps simply as ILC9 and
SMICA.)

Since we are interested in a rather diffuse structure, we apply a
low--pass filter by limiting the analysis to
multipoles $\ell \leq \ell_\text{max} = 20$ (as justified in
\S~\ref{sec:why_lmax_20}). Throughout, we work at a
\texttt{HEALPix}~\cite{Gorski:2004by} resolution of $\Nside=128$
(although the results are much the same at higher resolution because of our
$\ell_\text{max}=20$ filter). We compute the average temperature
$\overline{T}$ in a $\pm 2^\circ$ wide band around the Loop~I ring --- defined as 
a small circle of diameter $116^\circ$ centred at
\mbox{$(l, b) = (329^\circ,
  17.5^\circ)$}~\cite{Berkhuijsen:1971aa}.\footnote{The
  shape of Loop~I may differ from a sphere south of
  the Galactic plane~\cite{Vidal:2014boa} but we stick to 
  the original prescription~\cite{Berkhuijsen:1971aa}, as adopted in our
  earlier work~\citep{Liu:2014mpa}, since otherwise there would be additional
  uncertainty due to a `look elsewhere' effect.}  To evaluate the chance
probability of an average temperature at least as high, we simulate
$10^4$ maps with the WMAP 9\,yr best--fit angular power
spectrum~\cite{Bennett:2012zja}. Both the fraction of simulations with
a more extreme average temperature, and the $p$--value of a Gaussian
with the mean and standard deviation of the distribution of simulated
$\overline{T}$ are determined. We also investigate the effect of
applying the standard KQ85 and SMICA masks,\footnote{We had not used a
  mask earlier~\cite{Liu:2014mpa} because the WMAP Collaboration
  vouches for the absence of Galactic large--scale foregrounds in the
  ILC map ~\cite{LAMBDA}: {\it ``On angular scales greater than
    $\sim$10 degrees, we believe that the nine--year ILC map provides
    a reliable estimate of the CMB signal, with negligible instrument
    noise, over the full sky''}.} having downgraded these to
$\Nside=128$.

\begin{table}[!h]
\centering
\begin{tabular}{l l | r r r}
\hline
map & mask & $\overline{T} [\mu\text{K}]$ & $p$--value (from SD) & $p$--value (more extreme $\overline{T}$) \\
\hline
ILC9		& none	& 23.9		& 0.01		& 0.01 \\
ILC9		& KQ85	& 15.4		& 0.09		& 0.09 \\
SMICA	& none	& 22.8		& 0.01		& 0.01 \\
SMICA	& SMICA	& 21.8		& 0.02		& 0.02 \\
\hline
\end{tabular}
\caption{The mean temperature around Loop~I (and the chance
  probability) in various CMB maps.}
\label{tbl:Tmean}
\end{table}

The results are reported in Table~\ref{tbl:Tmean}. We find an
anomalously high average temperature of around 20~$\mu$K,  for both
the ILC9 and SMICA maps. The $p$--values are of order $1\%$, when not
applying a mask. For the SMICA map, the $p$--value increases to 2\%
when the SMICA mask is applied, but the average temperature remains
virtually unchanged ($\overline{T}=21.8$ versus
$22.8\mu\text{K}$). The increase in the $p$--value is therefore in
part due to the increased sample variance when applying the mask, and
not solely to masking of residual contamination from the Galactic
plane. The agreement between the $p$--values estimated in two separate
ways is just as expected for a Gaussian random field.

\section{Cross-correlation between temperature and pixel-to-loop distance}
\label{sec:corr}

Fig.~3 of our paper~\citep{Liu:2014mpa} shows that as the pixel
temperature increases, the angular distance between the corresponding
pixel and the centre of Loop~I decreases, which is a \emph{global}
effect and not just determined by the 4 highest temperature bins. This
can be quantified by the cross-correlation between $G(\text{p}_j)$ and
$T(\text{p}_j)$, where $G(\text{p}_j)$ is the angular distance (along
great circles) between the $j$-th pixel and the centre of Loop~I, and
$T(\text{p}_j)$ is the temperature of the $j$-th pixel---both computed
with \emph{no} binning. This cross-correlation coefficient is:
\begin{equation}
\label{equ:corr}
 C (G,T) = \frac{\sum{(G-\mu_G)(T-\mu_T)}}{\sqrt{\sum{(G-\mu_G)^2}\sum{(T-\mu_T)^2}}},
\end{equation}
where $\mu_T$ and $\mu_G$ are the sample average of $T$ and $G$ in the
Loop~I region. Using the low--pass filter with $\ell_\text{max}=20$ we
find $C(G,T)=-0.22$ for the ILC9 map and $C(G,T)=-0.20$ for the SMICA
map. As above, this result is insensitive to the map's resolution. In
addition, via its definition~(\ref{equ:corr}), the cross-correlation
is affected only by relative deviations from the mean, and not the
absolute amplitude. Both of these findings counter the criticism
~\cite{Ogburn:2014eca} of our work, viz. the alleged sensitivity to the map
pixelisation and the insensitivity of the method towards possibly
hotter peaks in the simulations.

\begin{figure}[t!]
\centering
\includegraphics[width=0.4\textwidth]{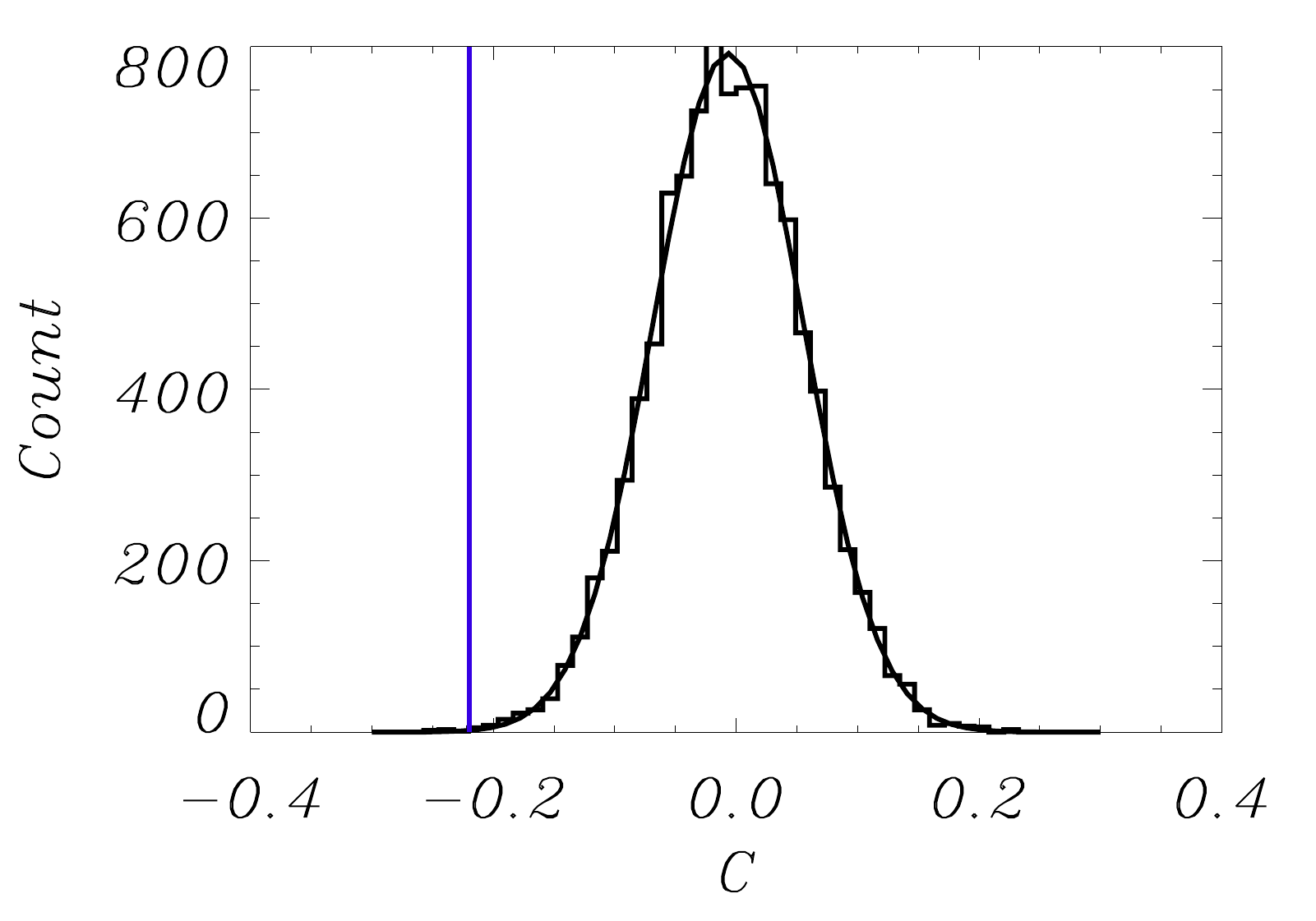}
\caption{The histogram of $C(G,T)$ for $10^4$ simulations. The blue
  line indicates the value obtained from~ILC9.}
\label{fig:corr hist}
\end{figure}

By generating $10^4$ simulations based on the best-fit $\Lambda$CDM
model power spectrum, we estimate the significance of finding
$C(G,T)=-0.22$ (for the ILC9 map) to be $\sim4 \times 10^{-4}$ and
$C(G,T)=-0.20$ (for the SMICA map) to be $\sim10^{-3}$. Again, the
distribution of $C(G,T)$ is well-fitted by a Gaussian (see
Fig.~\ref{fig:corr hist}) and both $p$--values are comparable to our
earlier result~\cite{Liu:2014mpa}. Hence our claim that there is a
$\sim24\mu$K anomalous signal from Loop~I which has somehow evaded
standard foreground cleaning techniques, does stand up to detailed scrutiny.

\section{Can the Loop~I anomaly be removed by masking?}
\label{sec:loop1 and mask}

It has been suggested~\cite{Ogburn:2014eca,Ade:2015qkp} that the
anomaly in the Loop~I region is not due to unsubtracted intrinsic
emission from Loop~I but rather contamination from other regions in
the Galactic plane along the line-of-sight. However if this were so,
one would not in general expect the hot spots inside the Galactic
plane mask to correlate with the Loop~I ring. If this happens by
chance, the anti--correlation ($C(G,T) < 0$) would be weaker when the
Galactic mask is applied. However we find that $C(G,T)$ is \emph{unaffected}
by the masks.

To check the effect of masking we reexamine the hot spots in the SMICA
map smoothed with $\ell_\text{max}=20$ using 4 different masks: SMICA,
Commander, NILC and SEVEM~\cite{Adam:2015tpy}. In Fig.~\ref{fig:masks
  and loop1} we mark the boundary of the masks with black lines and
indicate local maxima along Loop~I in the CMB temperature with black
circles. It can be seen that most of the local maxima lie
\emph{outside} the masks.  Unsurprisingly, the value of $C(G,T)$ is
nearly independent of the mask, or the resolution of the CMB map.
This is exhibited in Table~\ref{tab:td_corr3} along with the
corresponding significances from $10^4$ simulations.

\begin{figure}[h!]
\centering
\includegraphics[width=0.48\textwidth]{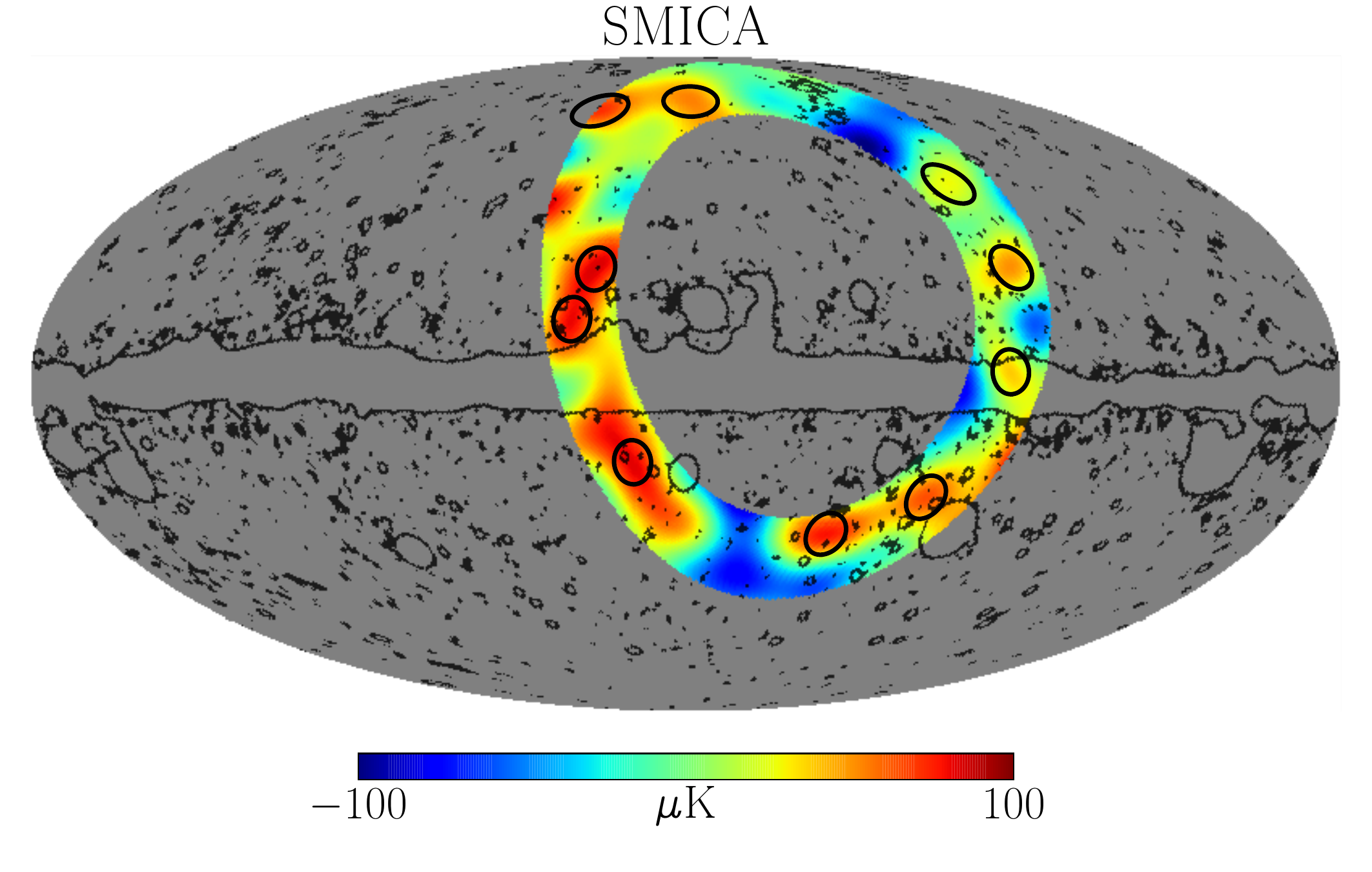}
\includegraphics[width=0.48\textwidth]{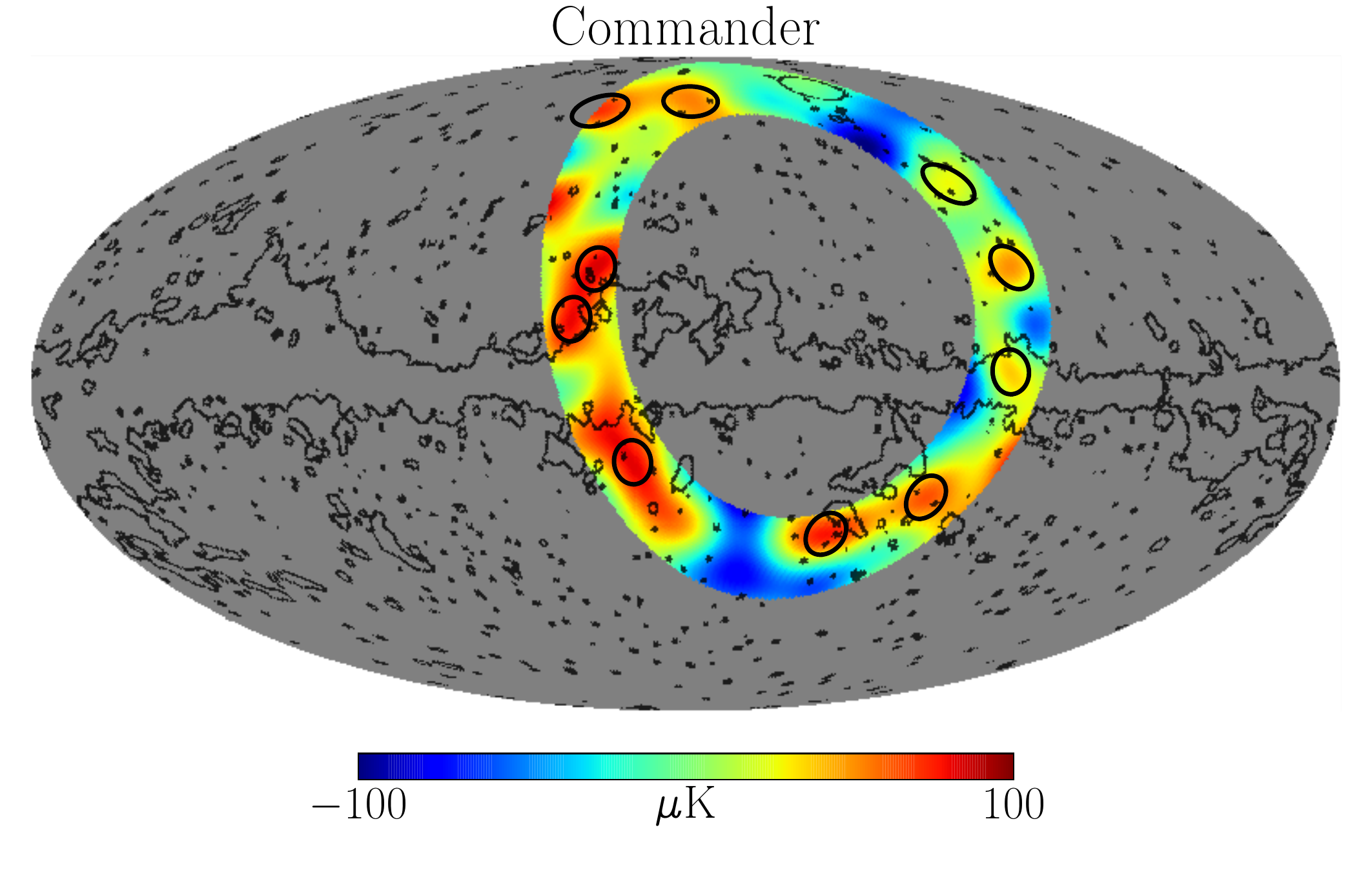}\\
\includegraphics[width=0.48\textwidth]{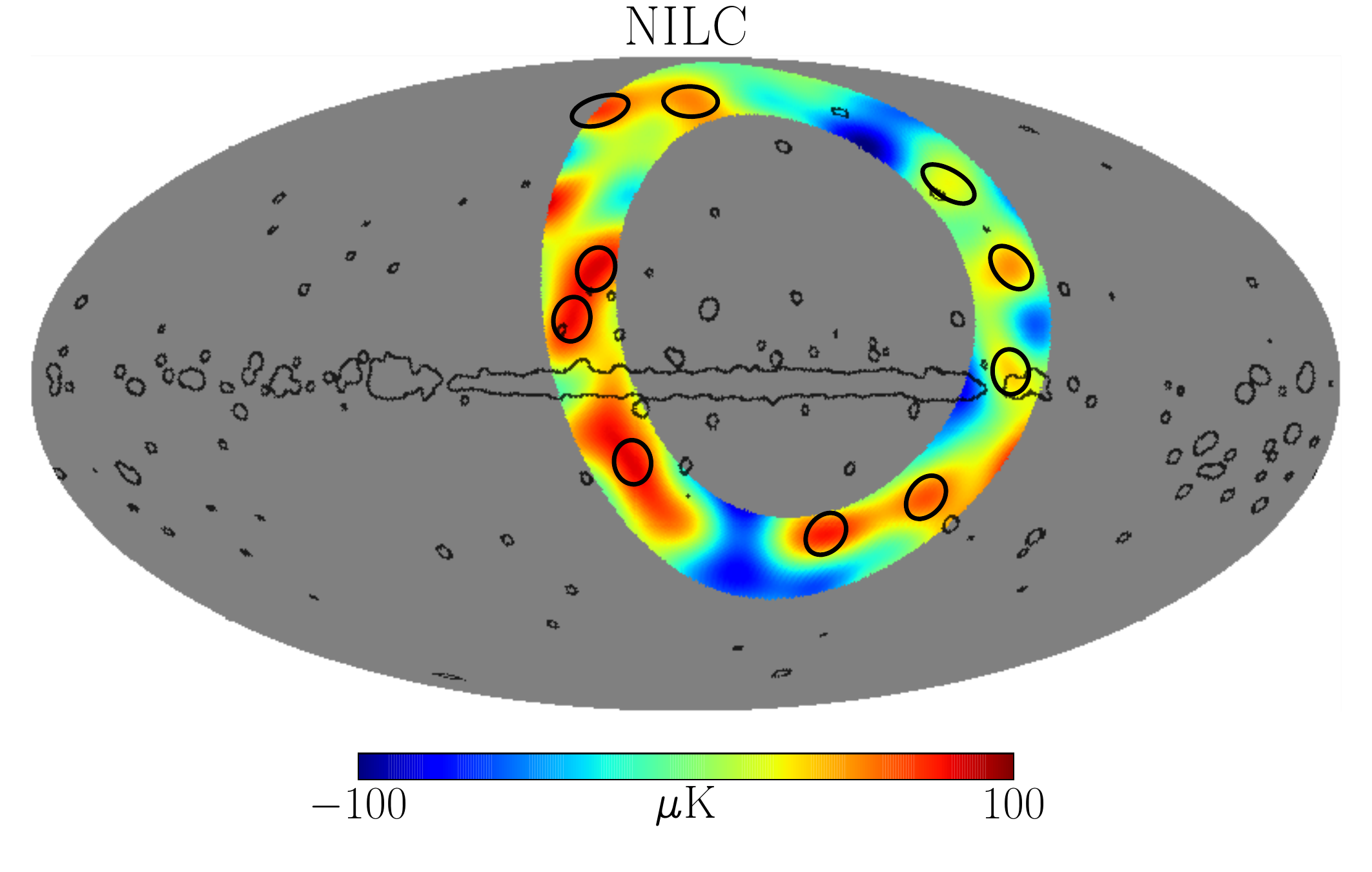}
\includegraphics[width=0.48\textwidth]{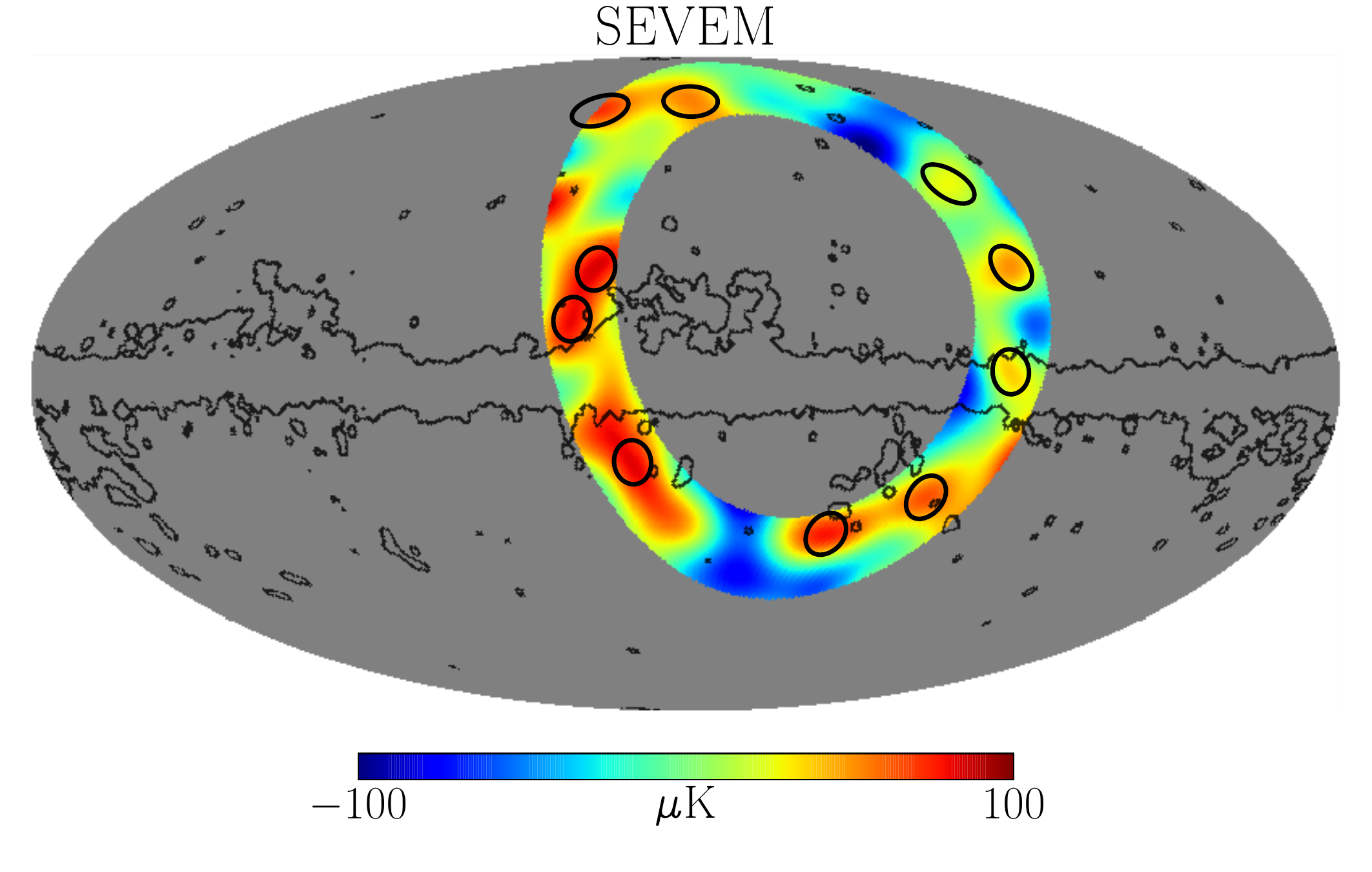}
\caption{The positions of local maxima of the temperature (black
  circles) along Loop~I relative to SMICA, Commander, NILC and SEVEM
  masks (black lines).}
\label{fig:masks and loop1}
\end{figure}

\begin{table}[t!]
\centering
\begin{tabular}{l l | r r r}
\hline
map & mask & $C(G,T)$ & $p$--value (from std.\ dev.) & $p$--value (more extreme $C(G,T)$) \\
\hline
ILC9		& none	& -0.22		& $3.9 \times 10^{-4}$	& $7.0 \times 10^{-4}$ \\
ILC9		& KQ85	& -0.17		& $1.4 \times 10^{-2}$	& $1.5 \times 10^{-2}$ \\
SMICA	& none	& -0.20		& $9.8 \times 10^{-4}$	& $1.4 \times 10^{-3}$ \\
SMICA	& SMICA	& -0.20		& $2.0 \times 10^{-3}$	& $2.6 \times 10^{-3}$ \\
SMICA	& Commander	& -0.20	& $2.8 \times 10^{-3}$	& $3.8 \times 10^{-3}$ \\
SMICA	& NILC	& -0.20		& $1.2 \times 10^{-3}$	& $1.7 \times 10^{-3}$ \\
SMICA	& SEVEM	& -0.20		& $2.1 \times 10^{-3}$	& $2.4 \times 10^{-3}$ \\
\hline
\end{tabular}

\caption{The temperature-angular distance correlation at
  $\Nside=128$. We quote the correlation coefficient $C(G,T)$, the $p$--value
  (assuming a Gaussian distribution of simulated $C(G,T)$'s) and the fraction of 
  simulations with smaller $C(G,T)$'s for different CMB maps and masks.}
\label{tab:td_corr3}
\end{table}

\section{Justification of the low-pass filter scale}
\label{sec:why_lmax_20}

\begin{figure}[h!]
\centering
\includegraphics[width=0.5\textwidth,clip=true,trim=0cm 0cm 0cm 0cm]{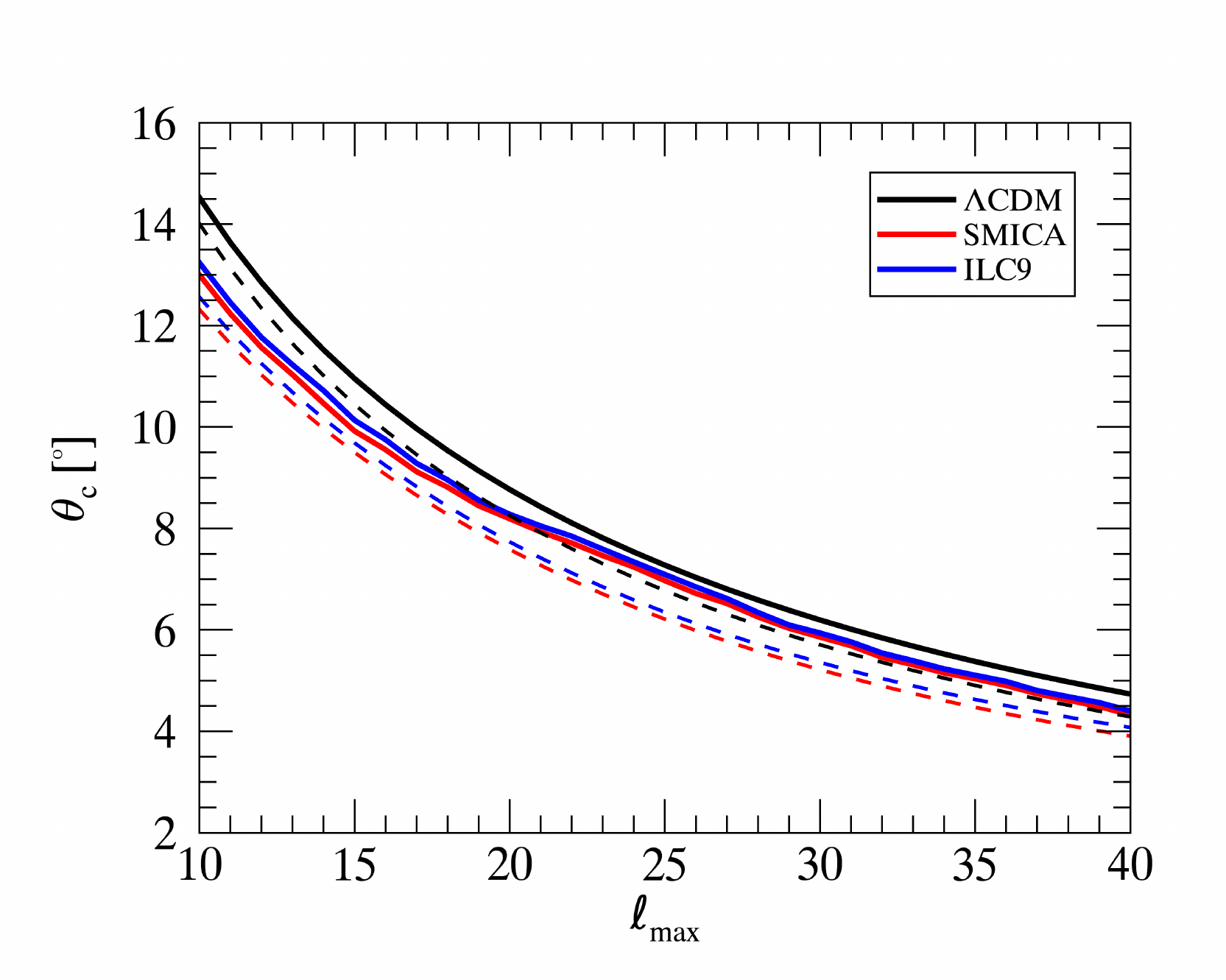}
\caption{Correlation angles $\theta_\text{c}$ (solid) and
  $\widetilde{\theta}_\text{c}$ (dashed) versus $\ell_\text{max}$, for the SMICA and ILC9 maps.}
\label{fig:dep}
\end{figure}

We now explain the choice of $\ell_\text{max}=20$ for filtering the
CMB map used  in our previous~\citep{Liu:2014mpa} as well as this work, by comparing
the angular width of Loop~I with the characteristic correlation angle
$\theta_\text{c}$ from the CMB maps. These ought to be comparable if we
wish to find a correlated signal between the two. The correlation
angle $\theta_\text{c}$ is defined using the two-point correlation
function $C(\theta)$ as:

\begin{equation}
C (\theta) 
\simeq C (0) + \frac{1}{2}C''(0)\theta^2 + \ldots \simeq C(0)\left(1-\frac{1}{2}\frac{\theta^2}{\theta_\text{c}^2}\right)
\label{eq:taylor}
\end{equation}
where the primes indicate differentiation with respect to
$\theta$. Hence
\begin{align}
\theta_\text{c}^2 \equiv -\frac{C(0)}{C''(0)}
= \frac{2\sum_{\ell=2}^{\ell_\text{max}}(2\ell+1)C_\ell}{\sum_{\ell=2}^{\ell_\text{max}}(2\ell+1)(\ell+1)\ell C_\ell} 
\label{corrangle}
\end{align}
clearly depends on the choice of
$\ell_\text{max}$. In addition to a sharp cutoff at $\ell_\text{max}$
we also consider a smooth cutoff
$\propto \exp(-\ell^2/\ell^2_\text{max})$ in the sums above and let
$\widetilde{\theta}_\text{c}$ denote the corresponding correlation
angle. We calculate these for different $\ell_\text{max}$ in the range
10--40 using the best-fit $\Lambda$CDM cosmology and power spectra
from the SMICA and ILC9 maps as shown in Fig.~\ref{fig:dep}. The
correlation angles for $\ell_\text{max}=20$ all lie around $8^\circ$
(see Table~\ref{tab:l}).

\begin{table}[h!]
\centering
\begin{tabular}{r|cc|cc}
\hline
 & \multicolumn{2}{c}{$\Nside=512$} & \multicolumn{2}{c}{$\Nside=16$} \\\hline
& \rule{0pt}{14pt}$\theta_c$ & $\widetilde{\theta_c}$ & $\theta_c$ &
                                                                     $\widetilde{\theta_c}$ \\\hline
$\Lambda\text{CDM}$ &	$8.77^\circ$ & $8.26^\circ$ & $8.77^\circ$ & $8.35^\circ$ \\
$\text{ILC9}$ & $8.28^\circ$ & $7.73^\circ$ & $8.46^\circ$ & $8.41^\circ$ \\
$\text{SMICA}$ & $8.19^\circ$	& $7.59^\circ$	& $8.33^\circ$ & $8.22^\circ$ \\
\end{tabular}
\caption{Values of $\theta_c$ and $\widetilde{\theta_c}$ for $\ell_\text{max}=20$ at $\Nside=512$ and $16$.}
\label{tab:l}
\end{table}

\begin{figure}[t!]
\centering
\begin{minipage}{0.45\textwidth}
\centering
\includegraphics[width=1\textwidth,clip=true,trim=2cm 7cm 2cm 8.5cm]{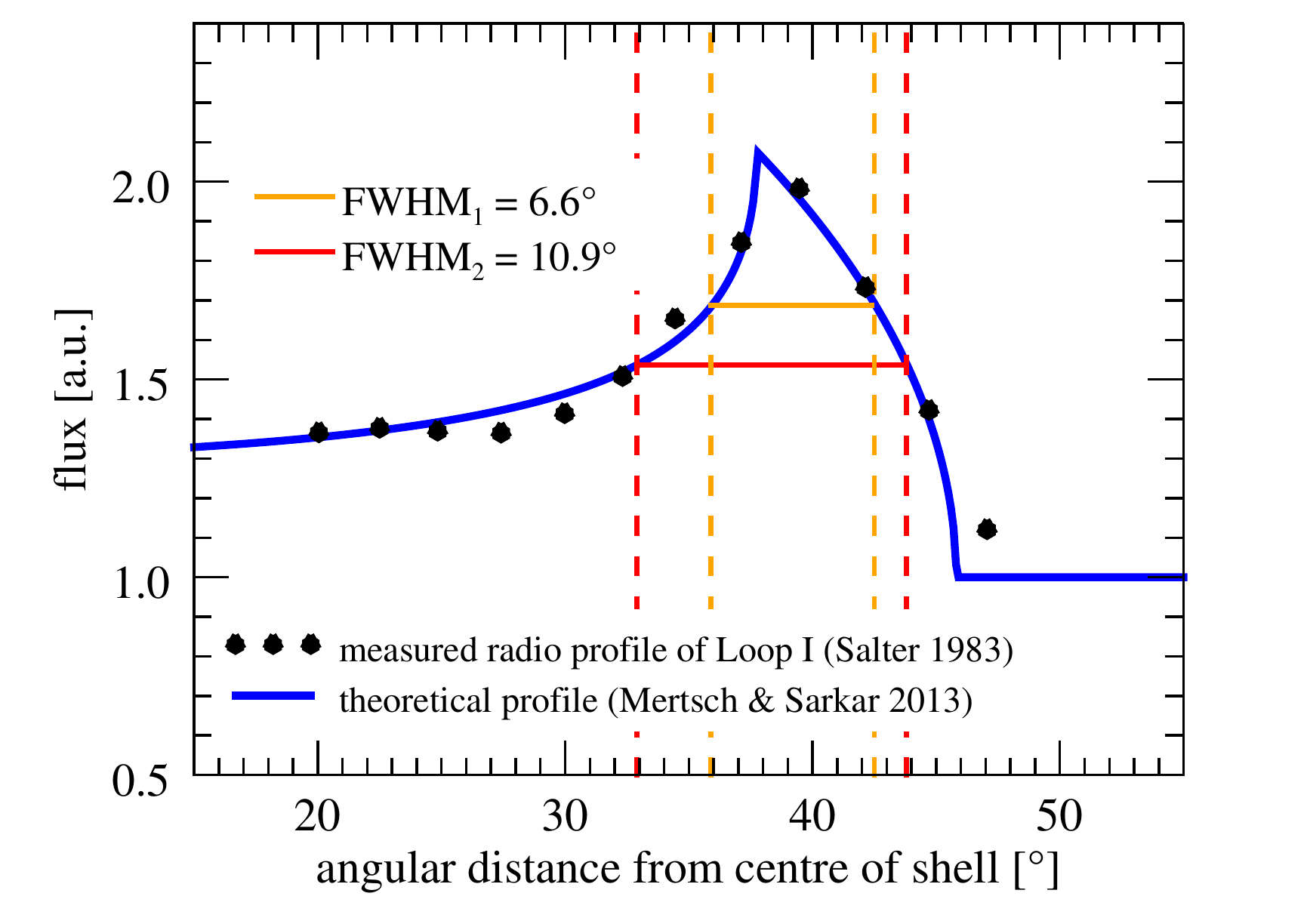}
\end{minipage}
\hfill
\begin{minipage}{0.49\textwidth}
\centering
\includegraphics[width=1\textwidth]{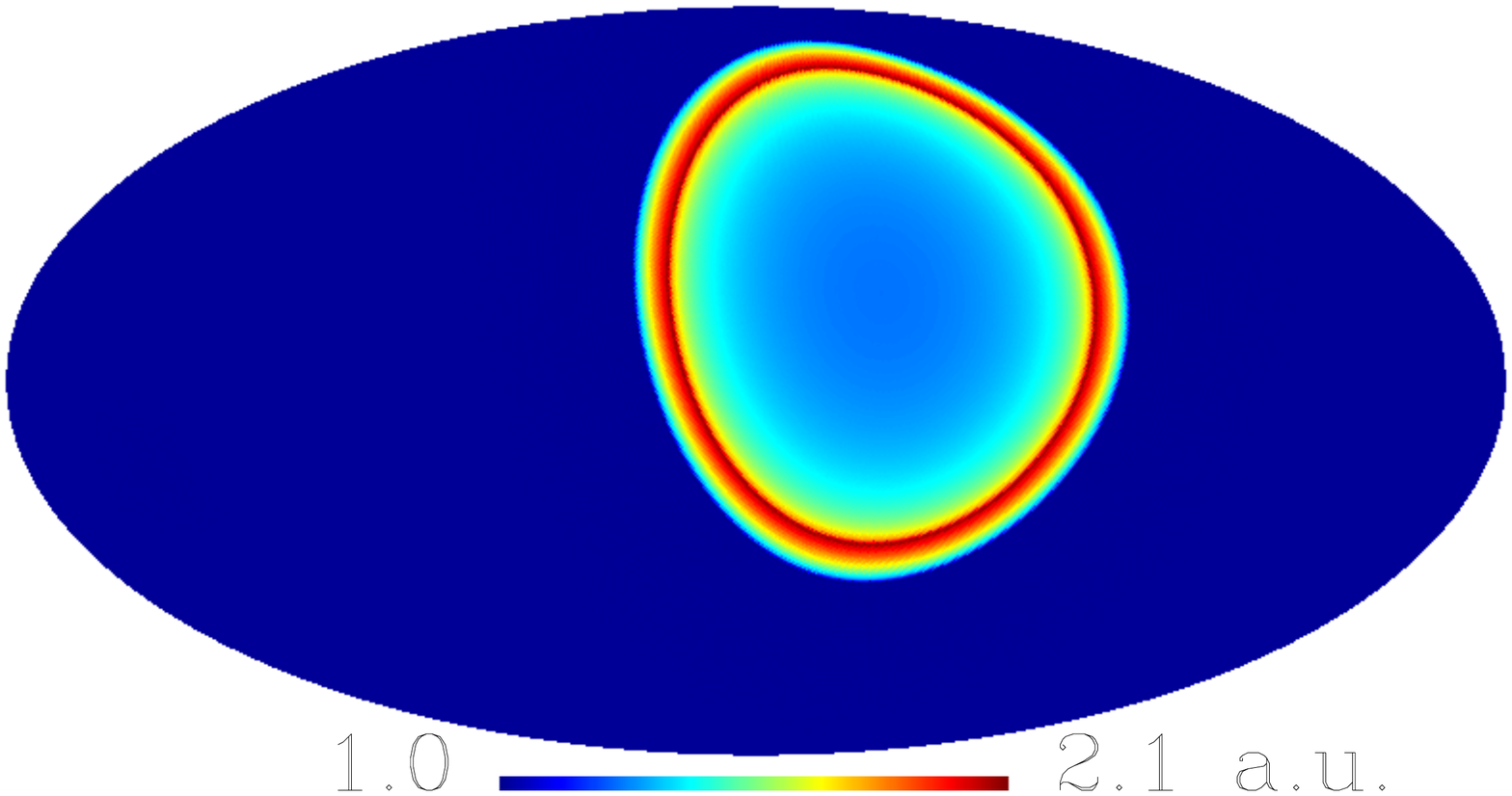}
\end{minipage}
\caption{The modelled angular profile of Loop I~\cite{Mertsch:2013pua}
  with the FWHM defined by two different measures (zero-level at
  centre or outside the loop) marked in orange and red, respectively,
  and its projection on the sky.}
\label{fig:profile}
\end{figure}

Observations~\cite{Salter:1983zz} of Loop~I are well-fitted by
modelling it as a shell of uniform emissivity~\cite{Mertsch:2013pua}
as seen in the left panel of Fig.~\ref{fig:profile}. We calculate the
full width at half maximum (FWHM) and find this to be
$\theta_{\text{FWHM}}=6.6^\circ$ if the zero-level is taken to be at
the centre. However there is emission above the background even here
--- had we chosen the zero-level outside of the loop the resulting
$\theta_{\text{FWHM}}$ would be~$10.9^\circ$, marked in red in the
left panel of Fig.~\ref{fig:profile}. The angular width of Loop~I is
thus comparable to the correlation angle for $\ell_\text{max}=20$,
justifying this as our choice~\citep{Liu:2014mpa} for smoothing the
CMB maps.

The power spectra used above were obtained from maps using
$\Nside=512$. We expect these results to hold for lower resolutions as
well, as the angular power does not change much below the value for
$\ell_\text{max}$ considered here. To illustrate this, we show in the
same Table~\ref{tab:l} values obtained from maps at $\Nside=16$ which
differ very little from the values obtained at higher resolution.

\section{Should anomalous emission be seen from other loops?}

The presence of a foreground residual that is morphologically
correlated with a well known Galactic structure, viz. Loop~I,
demonstrates the limitations of conventional (frequency--based)
foreground cleaning techniques. We had sugested~\cite{Liu:2014mpa}
that this residual is harder than synchrotron but also softer than
electric dipole emission from thermal dust, so may have a near
blackbody spectrum as expected of magnetic dipole emission from
magnetised dust~\cite{Draine:2012zu}. Such a previously unrecognised
(see however~\cite{Draine:2012wr}) foreground would understandably
pose a serious challenge for separating the CMB signal---in the
absence of any tracer map of MDE.

\begin{figure}[t!]
\centering
\includegraphics[width=0.48\textwidth]{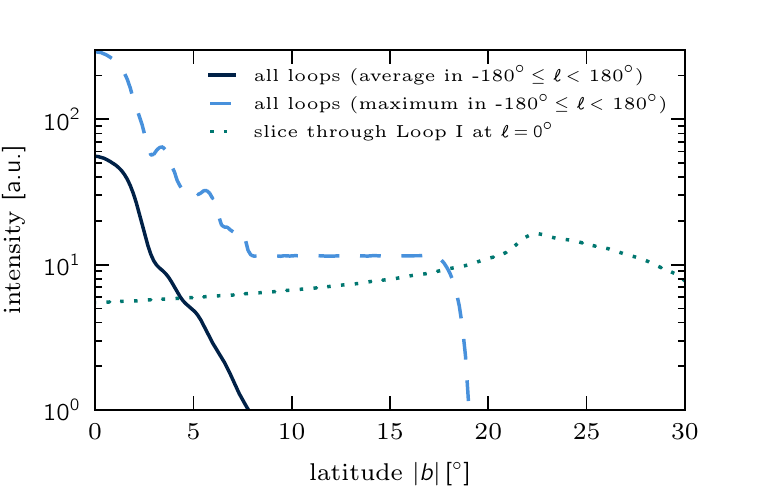}
\includegraphics[width=0.48\textwidth]{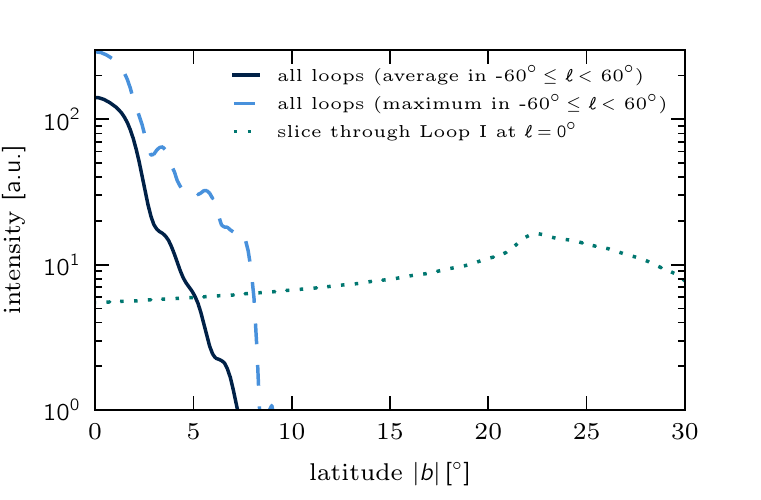}
\caption{Latitudinal profiles of the column density through the
  Galactic population of all loops in the longitude intervals
  indicated, compared to a slice through Loop~I.}
\label{fig:loop_profiles}
\end{figure}

As mentioned, the Planck Collaboration~\citep{Ade:2015qkp} have
argued that if MDE is the cause of the contamination in the Loop~I
region, such emission should have been observed from the Galactic disk
which harbours thousands of such old SNRs (`all loops')
\cite{Mertsch:2013pua}. We have checked this by using our simulation
of radio emission from the 4 known radio loops (including Loop~I) and
the Galactic population of old SNRs, using the same simulated maps as
shown in the middle row of Fig.~11 of Ref.~\cite{Mertsch:2013pua}. In
Fig.~\ref{fig:loop_profiles} we show the latitudinal profiles of the
column density through the Galactic population of all loops (both
the average and the maximum) in the longitude intervals
$[-180^{\circ}, 180^{\circ}]$ (left panel) and
$[-60^{\circ}, 60^{\circ}]$ (right panel). These are compared with a
slice through Loop~I at longitude $=0^{\circ}$.

It is seen that up to $|b| \sim 5^{\circ}$, the average contribution
from all Galactic loops \emph{dominates} over the contribution from
Loop~I. At higher latitude, however, the intensity of Loop~I is higher
and peaks at $b=-22^{\circ}$ (for the $l = 0^{\circ}$ slice). The
maximum of all Galactic loops in the longitude interval
$[-180^{\circ},180^{\circ}]$ is almost constant for latitude
$(8-18)^{\circ}$ where sight lines intersect with at most one loop. At
larger distances, the line of sight does not cross any Galactic loops
due to the excision in the simulation of heliocentric distances $<500$
pc (in order to avoid double
counting of local shells). For the longitude interval $[-60^{\circ}, 60^{\circ}]$ this drop
happens at even smaller latitude; due to the position of the Solar
system with respect to the nearest spiral arm, the closest (and
therefore spatially most extended) Galactic loops are in the second
and third Galactic quadrant (i.e. outside the interval
$[-60^{\circ}, 60^{\circ}]$).

In this analysis we have likely overestimated the column--depth for
the Galactic loops (or underestimated that of Loop~I). This is because
we have worked directly from the simulated \emph{synchrotron} maps and
the synchrotron emissivity of the loops is higher in the inner Galaxy
where cosmic ray electron density and the magnetic field strength are
larger. Dust emission, on the other hand, does not depend on these
variables hence the intensity from all loops close to the disk is
likely smaller than shown in Fig.~\ref{fig:loop_profiles}, making the
band where it dominates over Loop~I even narrower.

Thus we can conservatively conclude that the column density through
Galactic loops dominates over that of Loop~I only in a narrow band
around the Galactic plane, $|b| \lesssim 5^{\circ}$ (possibly even
smaller). It is our understanding that the presence of MDE in this
region cannot be excluded since component separation methods are least
effective here (cf. the Galactic masks shown in Fig.~\ref{fig:masks
  and loop1}). Moreover in the disk this emission necessarily
correlates with free--free emission: the distribution of SNRs closely
follows the free electron density and therefore the map of the average
column depth through old loops should be similar to the free--free
map. Given this morphological similarity, in supposedly
foreground--cleaned CMB maps like ILC or SMICA, part of the MDE may
have been misidentified as free--free emission, and perhaps even
labelled as the `anomalous microwave emission' which is seen to 
correlate with free--free emission~\cite{Ade:2015qkp}. We shall present 
a study of such possibilities elsewhere.

\section{Summary}

We have revisited the issue of a $\sim 20\mu\text{K}$ excess in the
Loop~I region discovered~\cite{Liu:2014mpa} in the WMAP ILC9 map and
have detected its presence in the 2015 Planck SMICA map as well. We
have presented an improved test statistic for investigating the
clustering of hot spots around Loop~I, confirming such clustering for
the ILC9 and SMICA map of the same order as was originally claimed. We
have found that even when the usual temperature analysis masks are
applied the anomaly still persists; for the SMICA map all four
analysis masks (SMICA, Commander, NILC and SEVEM) lead to $p$--values
of ${\cal O}(10^{-3})$. Additionally we provide a justification of the
low-pass filter scale ($\ell_\text{max}=20$) used, based on the
observed width of the radio profile of Loop~I. Finally, we show that
the similar anomalous emission that might be expected from other old
supernova remnants in the Galaxy is confined to a narrow band along
the Galactic plane and hence may not have been identified as
such. Polarisation maps from Planck will be essential to make further progress in
understanding this important foreground for the CMB, especially at
high galactic latitude.

\section*{Acknowledgements}

We acknowledge use of the
\texttt{HEALPix} package \cite{Gorski:2004by} and the WMAP
\cite{LAMBDA} and Planck \cite{planck} data archives. We 
are grateful to all participants from the Planck collaboration
in the 1-4 October 2015 meeting at Torun for inviting us to engage in
constructive discussions, in particular Anthony Banday, Hans-Kristian
Eriksen, Krzysztof Gorski, Patrick Leahy, Pavel Naselsky and Jean-Loup Puget.

H.L. is supported by the National Natural Science Foundation of China
(Grant No. 11033003), the National Natural Science Foundation for
Young Scientists of China (Grant No. 11203024) and the Youth
Innovation Promotion Association, CAS. P.M. is supported by DoE
contract DE-AC02-76SF00515 and a KIPAC Kavli Fellowship.
S.S. acknowledges a DNRF Niels Bohr Professorship.


\end{document}